\documentclass[graybox]{svmult}

\usepackage{type1cm} 
\usepackage{makeidx} 
\usepackage{graphicx} 
\usepackage{graphics}
\usepackage{subfig}
\usepackage{multicol} 
\usepackage[bottom]{footmisc}
\usepackage{newtxtext} %
\usepackage[varvw]{newtxmath} 
\usepackage[numbered]{matlab-prettifier}
\usepackage{microtype}
\usepackage{xcolor}

\makeindex 

\begin{document}
\title*{Statistical Tools for Frequency Response Functions from Posture Control Experiments: Estimation of Probability of a Sample and Comparison Between Groups of Unpaired Samples}
\titlerunning{Statistical Tools for FRFs}
\author{Vittorio Lippi\orcidID{0000-0001-5520-8974}}
\institute{Vittorio Lippi \at Institut f\"{u}r Digitalisierung in der Medizin, University of Freiburg, Germany;
Neurozentrum der Uniklinik Freiburg, University of Freiburg, Germany \email{vittorio.lippi@uniklinik-freiburg.de}}
%
%
\maketitle

\abstract{The frequency response function (FRF) is an established way to describe the outcome of experiments in posture control literature. The FRF is an empirical transfer function between an input stimulus and the induced body segment sway profile, represented as a vector of complex values associated with a vector of frequencies. Having obtained an FRF from a trial with a subject, it can be useful to quantify the likelihood it belongs to a certain population, e.g., to diagnose a condition or to evaluate the human likeliness of a humanoid robot or a wearable device. In this work, a recently proposed method for FRF statistics based on confidence bands computed with bootstrap will be summarized, and, on its basis, possible ways to quantify the likelihood of FRFs belonging to a given set will be proposed. Furthermore, a statistical test to compare groups of unpaired samples is presented.}

\section{Introduction}
\subsection{Overview}
\begin{figure}[tb!]
\centering
\includegraphics[width=1.00\textwidth]{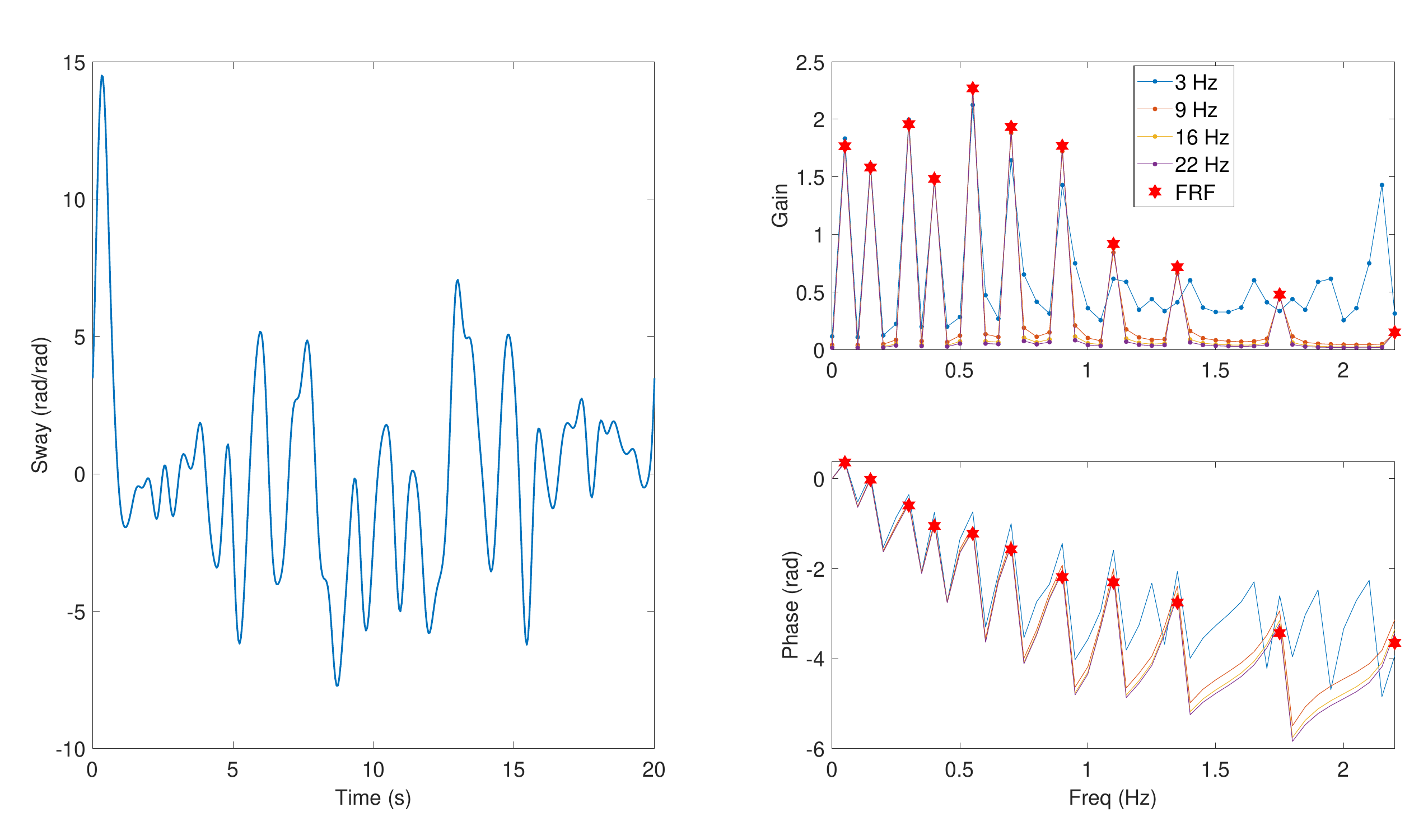}
\caption{The pseudo-impulsive-response (PIR) on the left, the FRF (red stars), and the DFT of the PIR (colored lines). The PIR is computed with eq. $x_i(t)= \sum^{M}_{k=1} \Re(H_{i,k}) \cos(2\pi \varphi_{k} t) + \Im(H_{i,k})\sin(2\pi \varphi_{k} t)$. The peculiarity of the FRF as proposed in posture control analysis is that the frequencies $\varphi=[ 0.05,\: 0.15,\: 0.3,\: 0.4, \:0.55,\: 0.7,\: 0.9,\: 1.1, \:1.35, \:1.75, \:2.2 ] Hz$ are not equally spaced. The associated $H$ is obtained as the average of the empirical transfer function over a range of frequencies. Such visualization of frequency response was defined for better plotting the signal \cite{peterka2002sensorimotor}. Different sample times are tested to reconstruct the FRF through a DFT. The period of the PIR is defined as the inverse of the greatest common divisor of the frequencies in $\varphi$; the sample time used in the examples is set to ten times the highest frequency in $\varphi$, i.e., 22 Hz. Notice how the gain tends to converge to zero between the peaks.}
\label{Pseudopulse}
\end{figure}
\textbf{The frequency response function} (FRF) is a common representation used in posture control experiments to describe the relationship between an input stimulus and the resulting body movement \cite{VANDIEEN2018219,1400711,hwang2016identification,amiri2019experimental,schut2019effect}. The FRF is defined as an empirical transfer function. The FRF is a complex function of frequency, and its structure must be considered when performing statistical analysis to assess differences between groups of FRFs. An example of FRF is shown in Fig.\ref{Pseudopulse} with a brief explanation. For an extended description of FRFs in posture control that goes beyond the limits of this paper, see \cite{peterka2002sensorimotor,lippi2020human}. The set used in the example is from \cite{lippi2020human}. It should be noted that FRF is a linear representation of the input-output relationship and that, in general, human posture responses are nonlinear. For example, they exhibit proportionally smaller responses (gains) to larger stimuli \cite{Hettich2013,hettich2015human,Hettich2014}. For this reason, when fitting linear models to posture control different parameters are obtained for different stimuli amplitudes, that is usually accounted as a \textit{reweighting} in the model \cite{asslander2014sensory,fb7822c9cc0445f5bf7a4339302a4ed5} due to different conditions (as observed for different availabilities of sensory inputs \cite{jeka2010dynamics,pasma2012sensory}). In the present work, the FRFs are regarded just as a description of the experimental trial, without any specific assumption on the underlying process; assumptions (i.e., linear or not linear model) can be part of a work using the presented methods.\\ 
\textbf{Statistics} are typically performed by defining a scalar variable to be studied, such as the norm of the difference between FRFs, or by considering the components independently. However, this approach can introduce an arbitrary metric that may have little connection with the experiment. To properly consider the nature of the FRF, a method oriented to complex functions should be used. One method, based on random field theory \cite{pataky2014vector}, considers the two components (imaginary and real) as independent variables \cite{Lippi2023}.\\
\textbf{The intuition} that an FRF, being a transfer function, can be transformed into a real-time domain signal without loss of information suggests an approach. On such real functions, the confidence bands can be defined using methods for continuous functions \cite{lenhoff1999bootstrap}. As the Fourier transform of a transfer function represents the impulsive response of a system, such function is referred to as \textit{pseudo-impulse-response}, PIR. The method to use bootstrap to define confidence bands on PIRs to perform statistics on FRFs is described in \cite{Lippiforthcoming}, and the code is available at \cite{LippiFRF24}.
The average PIR is
\begin{equation}
\bar{x}(t)= 1/N \sum_{i=1}^{N} x_i(t)
\end{equation}
and the STD is
\begin{equation}
\hat{\sigma}_x(t) = \sqrt{(1/(N-1))\sum^{N}_{i=1} \left|x_i(t)-\bar{x}(t) \right|^2}
\end{equation}
With these values, the prediction band can be defined for a new draw from the FRF distribution
$H_{n+1}$, and hence for the respective PIR $x_{N+1}(t)$.
With the desired confidence level $\alpha\%$, the constant $C_p$ is defined to obtain the probability
\begin{equation}
P\left[ \max\limits_{t} \left( \frac{|x_{N+1}(t) - \bar{x}(t)|}{\hat{\sigma}_{x}(t)} \right) \leq C_p\right] = \frac{\alpha}{100}
\label{confidenceeq}
\end{equation}
and the prediction band for a new FRF is
\begin{equation}
\bar{x}(t) \pm C_p \cdot \hat{\sigma}_{x}(t).
\end{equation}
This work proposes a further application of such bands to quantify the degree to which a sample FRF belongs to a distribution. The bootstrap approximates the probability in eq. \ref{confidenceeq} as:
\begin{equation}
\frac{1}{B}\sum\limits_{b=1}^{B}\left[\frac{1}{n}\sum\limits_{i=1}^{n}I\left(\max\limits_{t} \left( \frac{|x_i(t) - \hat{x}^b(t)|}{\hat{\sigma}^b_{x}(t)} \right) \leq C_p\right)\right]
\label{bootstrapped}
\end{equation}
{\color{black} where the function $I()$ is equal to $1$ when the condition in parentheses is verified and equal to $0$ otherwise. In practice, the inner sum is the number of samples inside the prediction band for the $b^{th}$ bootstrap iteration.} Eq. \ref{bootstrapped} is the average, over the $B$ bootstrap replications, of the proportion of the original data curves whose maximum standardized deviation from the bootstrap mean is less than or equal to $C_p$. The superscript $^b$ addressed that the quantity is computed based on the resampled set. Notice that $\hat{\sigma}^b_{x}(t)$ is based on the resample set at every iteration of the bootstrap as recommended in \cite{hall1991two}. Such ``pivotization'' also allows for null hypothesis testing without having to simulate the distribution produced by the null hypothesis \cite{davison2003recent}, as will be shown in the examples. The number $B$ is set to be relatively large with a trade-off between the accuracy and the computational time. Different indications about the required $B$ are discussed in \cite{zoubir1998bootstrap}.
In a \textit{opposite} situation to what is done in \cite{Lippiforthcoming}, the $C_p$ is known (inferred using the tested sample), and the $\alpha$ is the quantity to be computed. The statistics obtained from the bootstrap repetitions, i.e., the inner term from eq. $\left( \max\limits_{t} \left( \frac{|\hat{x}(t) - \hat{x}^b(t)|}{\hat{\sigma}^b_{x}(t)} \right) \right)$ are sorted in ascending order, and a histogram approximating the cumulative density function is computed. {\color{black} The desired $\alpha$ is at the intersection of the cumulative histogram with the threshold $C_p$, as shown in Fig.\ref{fig:TimeDomainMinPrediction}, right.}

\subsection{Probability that a sample belongs to a distribution.} 
\begin{figure}[htbp]
\centering
\includegraphics[width=1.00\textwidth]{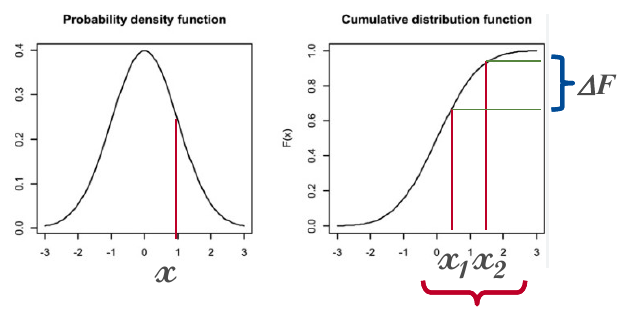}
\caption{Approximation of the pdf using an incremental ratio on the cdf estimated by the bootstrap. $\Delta F(x)$ is a fixed quantity (here $1/10$ of the number of samples $N$) and $x_1$ and $x_2$ are the values of $x$ found in the vector ordered distances $D$ produced by the Bootstrap moving back and forward of $N \cdot\Delta F(x)$ positions.}
\label{fig:CDF_PDF}
\end{figure}

The need to classify FRFs in groups arises when the posturography is used to diagnose a condition or to assess the human likeness of a behavior produced by a robot humanoid or wearable as in \cite{lippi2020human}. A modified version of the function computing the prediction band from \cite{Lippiforthcoming} can compute the minimal band, including a given sample. See fig. \ref{fig:TimeDomainMinPrediction}. This works by reversing the process used to compute the prediction band, i.e., finding the confidence $\alpha$ given the distance from the mean: the maximum distance between the test sample and the estimated mean in the histogram produced by the bootstrap as shown in Fig. \ref{fig:TimeDomainMinPrediction}.\\
\textbf{Approximation of the probability density function.} Further measures can be defined as the empirical estimation of the probability density function (PDF) and cumulative density function (CDF). As an example, in the present work, they are defined on the distance $D= \int (x_i(t)-\hat{x}(t))^2 dt$, but the principle can be generalized (the function in the library can take a generic function to be used as metric as input). The CDF $F(x)=P[X \leq x]$is computed empirically with a bootstrap (a mean and a STD are provided for the estimate), and the PDF is computed by approximating $f(x) = dF/dx\approx \Delta F(x)/(x_2-x_1) $ where $\Delta F(x)$ is a fixed quantity (here $1/10$ of the number of samples $N$) and $x_1$ and $x_2$ are the values of $x$ found in the vector ordered distances $D$ produced by the bootstrap moving back and forward of $N \cdot\Delta F(x)$ positions. The approach is exemplified in Fig \ref{fig:CDF_PDF}.
\begin{svgraybox}
The minimal prediction band and the estimated CDF indicate how much a sample is far away from the average of the distribution. The approximated pdf is an estimation of the probability that a sample is part of a distribution. These three measures can be used for diagnostics, e.g., comparing a subject with healthy and patient groups to test for pathological patterns, or evaluation, e.g., to measure the human likeness of an FRF produced by a robot compared to a sample of healthy humans. 
\end{svgraybox}

\subsection{Comparison between unpaired samples.} 
\begin{figure}[htbp]
\centering
\subfloat{\includegraphics[width=0.50\textwidth]{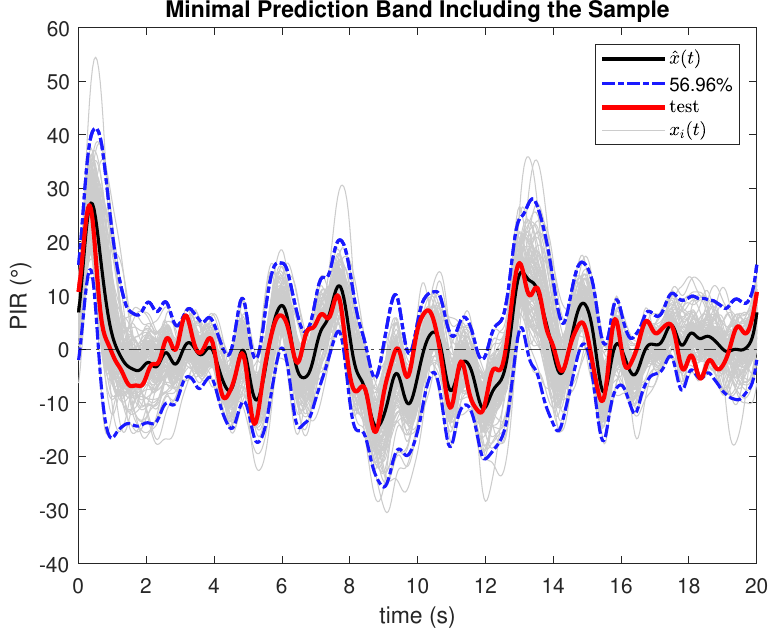}}
\subfloat{\includegraphics[width=0.50\textwidth]{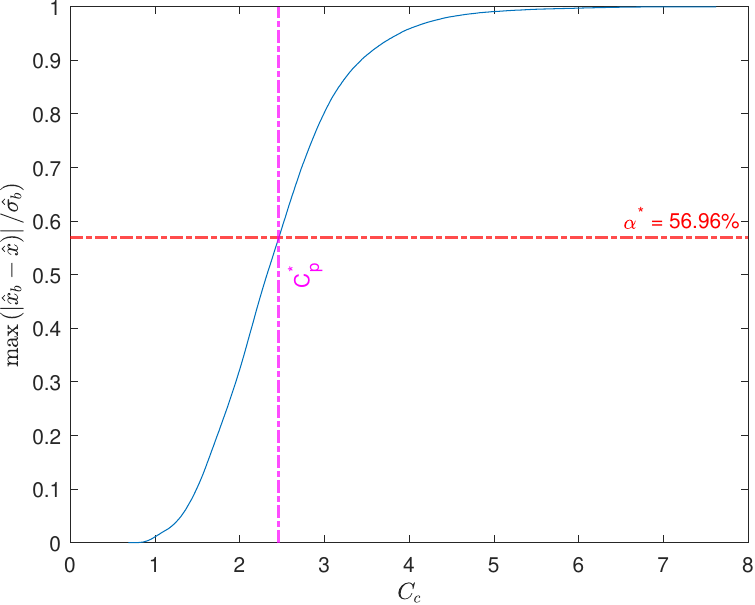}}
\caption{Minimal prediction band (left) and the cumulative histogram used to compute $\alpha$ (right). The histogram is produced by the bootstrap approximating the probability as $\frac{1}{B}\sum\limits_{b=1}^{B}\left[I \left( \max\limits_{t} \left( \frac{|\hat{x}(t) - \hat{x}^b(t)|}{\hat{\sigma}^b_{x}(t)} \right) \right) \leq C_c\right]$. In this case, opposite to what is done in \cite{Lippiforthcoming}, the $C_p$ is known (inferred using the tested sample), and the $\alpha$ is the quantity to be computed.}
\label{fig:TimeDomainMinPrediction}
\end{figure}

{\color{black} The confidence band on the difference between the groups' mean can can be computed, given the desired confidence level $\alpha$.} The function $p(t)$ considered is the difference between the averages of the two groups. This means that $N$ samples are produced with bootstrap repetitions. Given the desired confidence level $\alpha\%$ the constant $C_u$ is defined to obtain the probability:
\begin{equation}
P\left[ \max\limits_{t} \left( \frac{|p(t) - \hat{p}(t)|}{\hat{\sigma}_{p}(t)} \right) \leq C_u\right] = \frac{\alpha}{100}
\label{confidenceequnp}
\end{equation}
The $\alpha\%$ confidence band for $\hat{x}(t)$ is then
\begin{equation}
\label{confidenceconstantunp}
\hat{x}(t) \pm C_u \cdot \hat{\sigma}_{p}(t)
\end{equation}
\textbf{The bootstrap} is used to determine $C_u$ . Approximated versions of the probabilities in eq.\ref{confidenceequnp} are obtained using empirical distributions produced by resampling the sample set. The constant $C_u$ is set so that the approximated probability is as close as possible to the desired confidence $\alpha\%$. {\color{black} This is done using the histogram in a way similar to what was shown Fig.\ref{fig:TimeDomainMinPrediction} for the estimated probability of a sample belonging to a distribution, but in this case $\alpha$ is set a priori then $C_u$ is found at the intersection between $\alpha$ and the cumulative histogram.}
Specifically, eq. \ref{confidenceequnp} has the following bootstrap approximation:
\begin{equation}
\frac{1}{B}\sum\limits_{b=1}^{B} I \left[ \max\limits_{t} \left( \frac{|\hat{p}^b(t) - \bar{p}(t)|}{\hat{\sigma}^{b_{N}}_{p}(t)} \right) \leq C_u\right]
\label{confidencebootstrapunp}
\end{equation}
Eq. \ref{confidencebootstrapunp} is the average, over the $B$ bootstrap replications, of the proportion of the original data curves whose maximum standardized deviation from the bootstrap mean is less than or equal to $C_u$. The superscript $^b$ addressed that the quantity is computed based on the resampled set, and $^{b_{N}}$ means that the quantity is computed with a nested bootstrap loop. In fact, $\hat{\sigma}^{b_{N}}_{p}(t)$ is based on the resampled set at every iteration of the bootstrap.

\begin{figure}[htbp]
	\centering
		\includegraphics[width=1.00\textwidth]{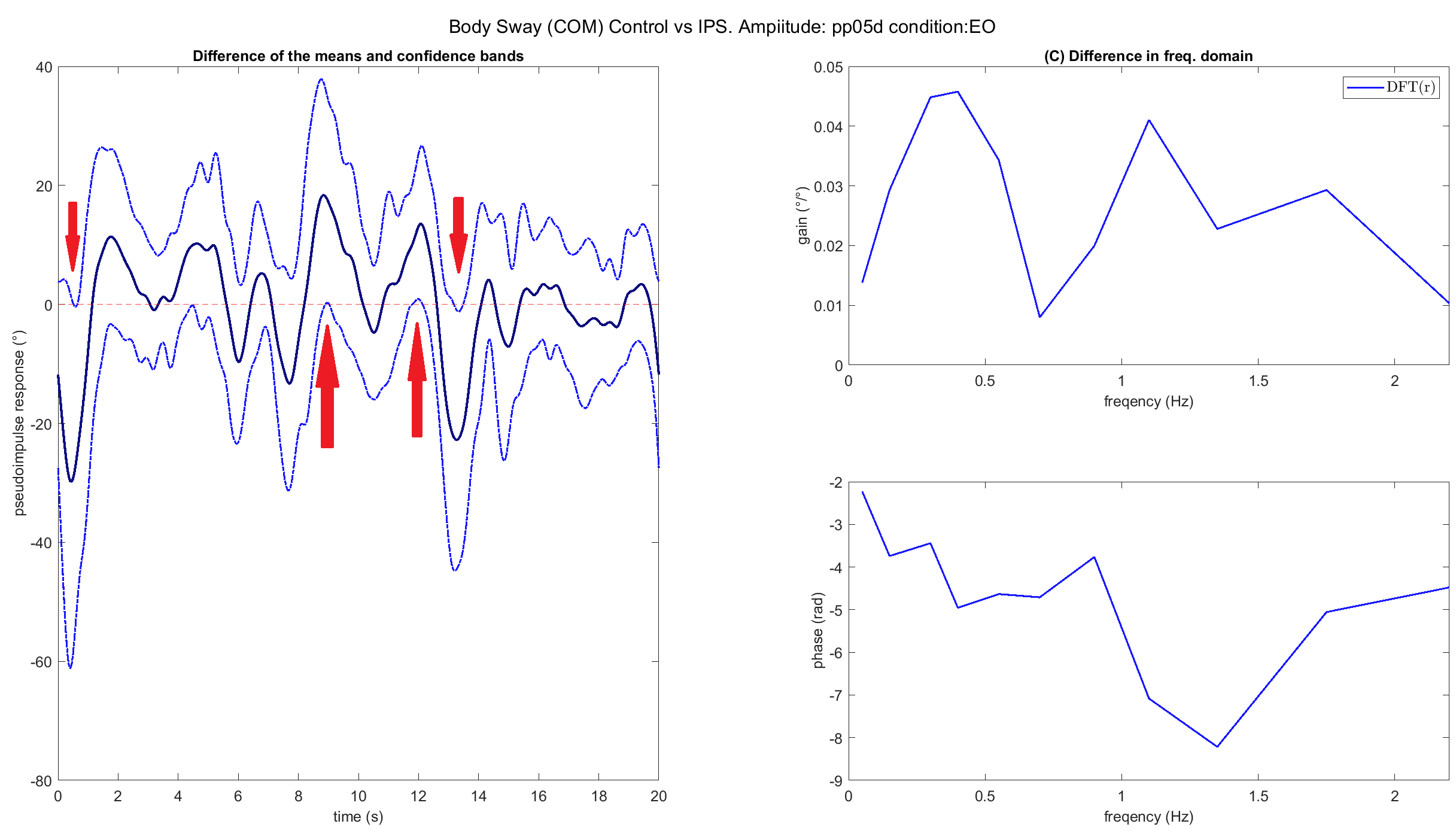}
	\caption{example of the use of the test for the comparison of unpaired groups. Head sway responses to support surface tilt of stage Idiopathic Parkinson's disease (IPD) and healthy control group (11 and 17 subjects respectively \cite{LippiSubmitted}). On the left, the average (over bootstrap repetitions) difference between the mean PIR (over the groups) is compared with the x-axis, i.e., PIR = 0, representing the null hypothesis that the averages of the two groups are the same. The $\alpha=95 \%$ confidence bands are represented with the dotted lines. The test result is the rejection of the null hypothesis because the x-axis is outside of the bands (see arrows). On the right, the residuals $r$ (band exceeding the x-axis) are transformed in the frequency domain with a discrete Fourier transform $DFT(r)$. This way, the groups' differences can be visualized and discussed in the same domain as the FRFs.}
	\label{UnpairedExample}
\end{figure}

\begin{svgraybox}
The test can be performed to compare two groups, as shown in Fig.\ref{UnpairedExample}.
The FRFs of the two groups and the desired $\alpha$ are the input of the function computing the confidence levels (\S\ref{UnpairedExample}). The result is an average difference between the groups and the associated confidence bands. The difference can be compared with the x-axis, i.e., zero difference according to the null hypothesis. If the x-axis is outside the confidence bands, the hypothesis is rejected with $p<1-\alpha$.
\end{svgraybox}

\subsection{ \color{black} State of the Art, Previous Work, and Original Contribution}
{\color{black} At the state of the art, there is no established method for performing statistics on FRFs for posture control.
Statistics on FRFs are often performed by defining a scalar variable, such as the norm of the difference between FRFs, or by analyzing components independently, applied to real and complex components separately \cite{lippi2023human,Akcay2021}. Sometimes, both approaches are combined, e.g., frequency-by-frequency comparison as a post-hoc test when the null hypothesis is rejected on the scalar value \cite{lippi2020body}. Multivariate methods like Hotelling's T2 are used for complex values, as in \cite{asslander2014sensory}, with further post-hoc tests applying bootstrap on magnitude and phase separately. Using a scalar variable introduces an arbitrary metric that may be justified when assumed a priori as in \cite{lippi2020human}, where a human-likeness score is defined. Testing frequencies or components separately overlooks the dependency among FRF values, and multiple comparison corrections like Bonferroni can overly reduce experimental power. A method focused on complex functions is needed. In \cite{Lippi2023}, a preliminary method based on random field theory \cite{pataky2016region} considers imaginary and real components as independent variables, using a 1-D Hotelling T2 test in the frequency domain \cite{pataky2014vector}.

Statistical analysis of continuous data using confidence bands is reviewed in \cite{joch2019inference}. The \textit{function-based resampling technique} (FBRT) fits data into mathematical functions, using coefficients to estimate variability and a scaling factor to adjust prediction coverage. Typically, sinusoidal functions with Fourier coefficients are used. However, selecting appropriate functions can be challenging. For example, while assuming a \textit{low-pass} power spectrum works for joint trajectories during gait \cite{lenhoff1999bootstrap}, biological signals like EEG lack specific representative frequencies across samples \cite{joch2019inference}.
}

The methods described in this paper represent the latest extension of the FRF-statistics library \cite{LippiFRF24} that originally included tools to define confidence and prediction intervals on FRFs as described in \cite{Lippiforthcoming}. {\color{black} Despite the library being publicly available, The present paper represents the first time the method is published}. The idea of estimating the probability of a sample FRF belonging to a distribution was presented in a conference talk \cite{Lippi2024a}, of which the present paper represents an extended version. While the confidence bands presented in \cite{Lippiforthcoming} can be used to test the difference between groups with paired samples (i.e., studying the mean of the difference between the couples of paired samples), a specific function must be implemented for the unpaired test. Such function has been used in a recently submitted work \cite{LippiSubmitted} but has yet to be published with its details, and hence, it represents an original contribution of the present paper.

\section{Materials and Methods}
\textbf{The estimated PDF and minimal prediction bands} are tested using data from previous experiments \cite{lippi2020human,robovis21}. In particular, the FRFs represent the body sway response to support surface tilt. 

\begin{figure}[htbp]
	\centering
		\includegraphics[width=1.00\textwidth]{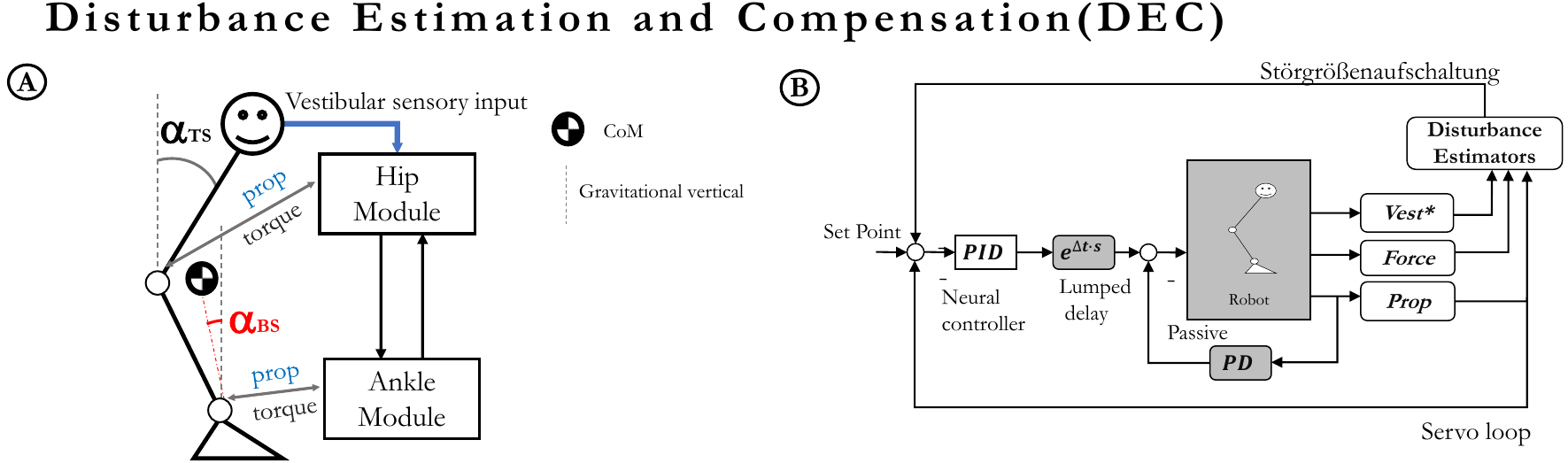}
	\caption{Overview of the control system.  A: The DEC, implemented as a double inverted pendulum (DIP) \cite{Hettich2013,hettich2015human}, with two control modules: the ``Hip Module'' controlling the orientation of the trunk in space $\alpha_{TS}$, and the ``Ankle Module'' controlling the orientation of the body CoM in space $\alpha_{BS}$. B: The scheme of a module of the DEC control. The system includes a passive control loop (PD) and a neural controller (PID) implementing the servo loop and the compensation of estimated disturbances as a \textit{st{\"o}rgr{\"o}{\ss}enaufschaltung}, i.e., feedforward compensation based on sensory input \cite{bleisteiner2013handbuch}. Each module has a lumped delay $e^{\Delta t \cdot s}$. The ankle module's vestibular input \textit{vest*} uses a sensor fusion-derived information (vestibular + proprioceptive) to reconstruct the orientation of legs in space. The vestibular input directly provides the head's orientation in space (and hence of the trunk in space).}
	\label{fig:ControlSystems}
\end{figure}

\textbf{The comparison between groups of unpaired samples} are tested with simulated data. Specifically, a double inverted pendulum description of upright posture control has been implemented using the bio-inspired model DEC, i.e. disturbance estimation and compensation \cite{lippi2017human,lippi2016human}, resulting in a model with two control modules controlling two degrees of freedom in the sagittal plane, as presented in \cite{Hettich2013,hettich2015human,zebenay2015human}. The model is shown in Fig. \ref{fig:ControlSystems}. 

\begin{figure}[htbp]
	\centering
		\includegraphics[width=1.00\textwidth]{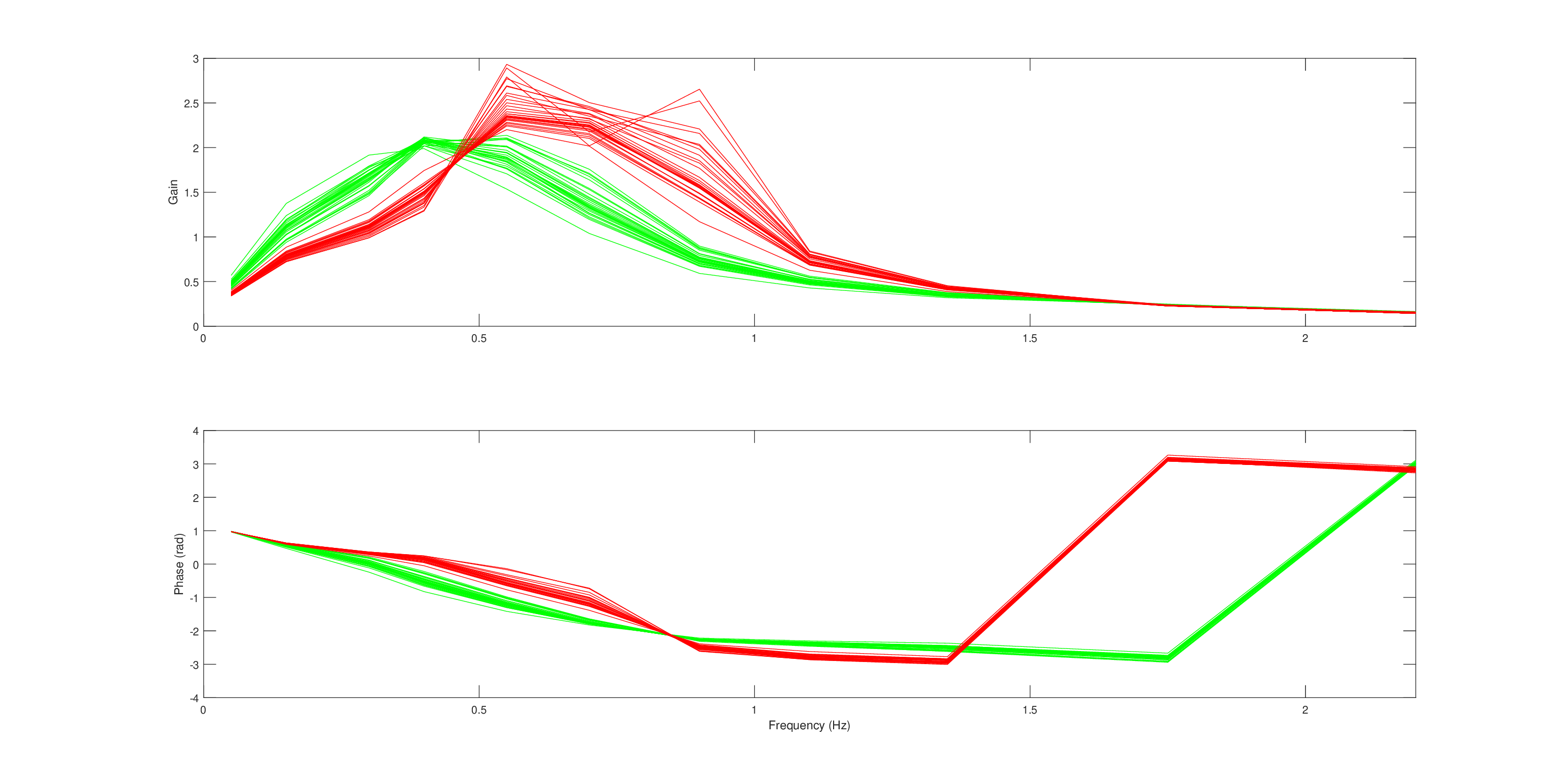}
	\caption{Responses simulated with a double inverted pendulum model perturbed with support surface tilt (PRTS, peak-to-peak $1^{\circ}$)}
	\label{fig:SimulatedUnpaired}
\end{figure}

A Base set of parameters from previous works \cite{Hettich2013,hettich2015human} have been used. The parameters have been altered to simulate two groups of 30 individuals by modifying the stiffness of the hip and ankle joints. This scenario can represent differences between young and elderly subjects as reported in \cite{5247102}. Specifically, an increase of $60\%$ in the stiffness was used to differentiate the two groups, and a normally distributed variable with STD equal to $10\%$ of the stiffness was used to differentiate between subjects. The input support surface tilt is a pseudo-random ternary signal (PRTS) profile with peak-to-peak amplitude of $1^{\circ}$. The resulting FRFs describing the relationship between support surface tilt are shown in Fig. \ref{fig:SimulatedUnpaired}. {\color{black} The use of the PRTS as input was presented in \cite{peterka2002sensorimotor}. Such a signal presents the advantage that it is not predictable by human subjects \cite{peterka2004dynamic}. Its power spectrum has a \textit{comb-shaped} profile, i.e., peaks alternated to zeroes that are reflected in the FRF (Fig.\ref{Pseudopulse}). In the general case, the method presented here can potentially be applied to the tests with other stimulus profiles, e.g., the sum of sinusoids (SoS) \cite{MAKI19931181,10.3389/fnhum.2024.1471132}, or even impulsive responses \cite{monteleone2023method},  although in that case there is not a predefined set of frequency to be selected.}

\section{Results}
\subsection{Minimal Prediction Band and Estimated PDF on Human Experiment Data}
To show how the measures work, a sample has been removed from the set (Sway responses to platform tilt from \cite{lippi2020human,robovis21}), and its minimum prediction band has been computed based on the rest of the samples, as shown in Fig. \ref{fig:TimeDomainMinPrediction}. The result was $\alpha=56.98\%$ The estimated probability distribution features were: cdf ($F = 0.6075$, $\sigma_F = 0.0523$) and pdf ($f = 0.0042$, $\sigma_f = 0.0016$).
\subsection{Comparison with Unpaired Samples}
\begin{figure}
	\centering
		\includegraphics[width=1.00\textwidth]{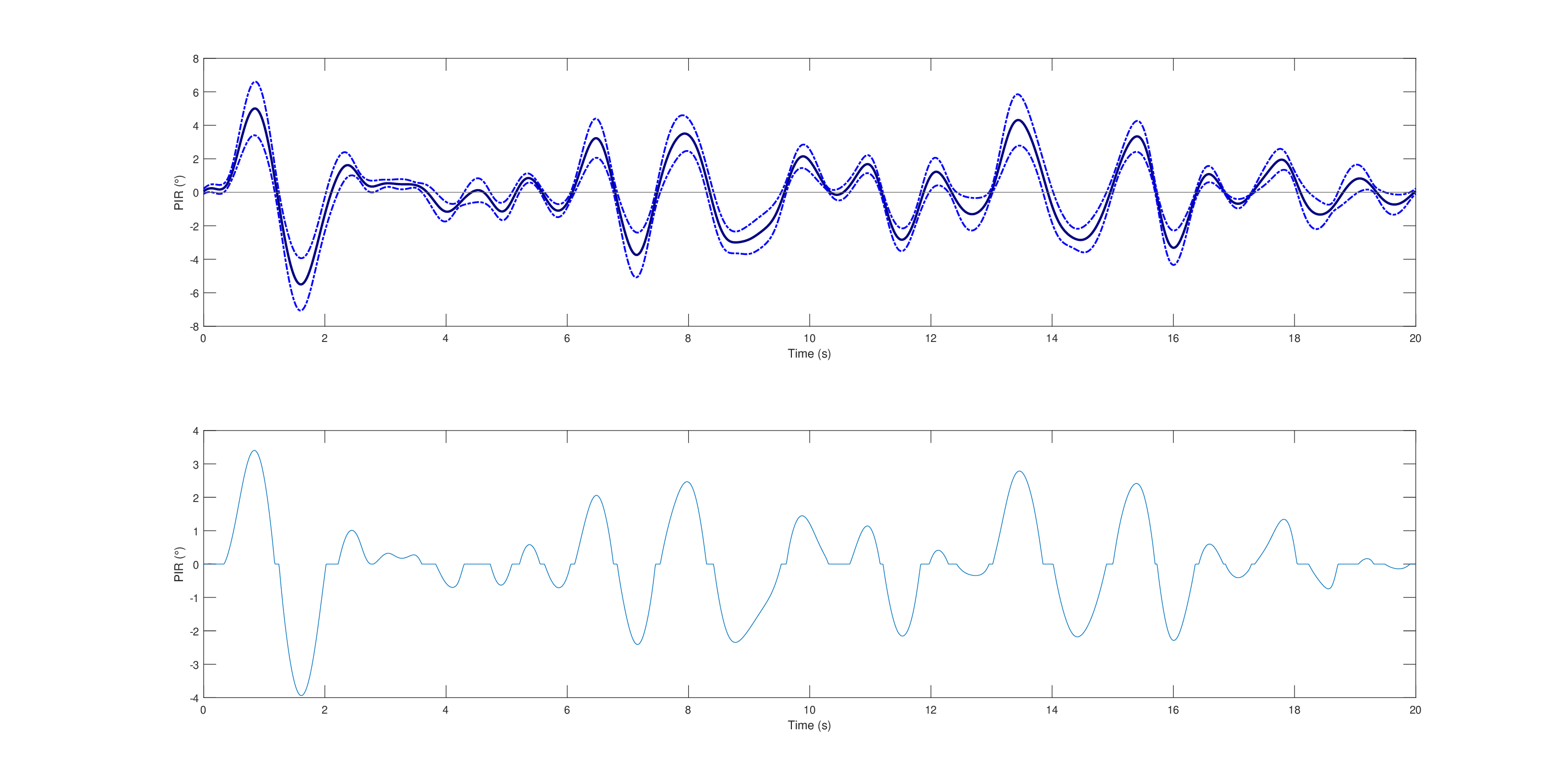}
	\caption{Average difference between the means of the two groups of PIRs and residuals. in the plot above, the average of the difference across the bootstrap repetition (solid blue line) is plotted with the $95\%$ confidence bands (eq. \ref{confidenceconstantunp}). the horizontal line $x=0$ represents the null hypothesis that the difference between the means is zero. As such a line is outside the confidence intervals, the difference between the two groups is considered significant with $95\%$ confidence. In the plot below, the residuals, i.e., the difference between the confidence bands and the horizontal axis, provide a visualization of where and how the difference between the means exceeds the threshold of the confidence levels. The Fourier transform of the residuals is shown in Fig. \ref{fig:residualFRF} to localize the difference in the frequency domain.}
	\label{fig:SimulatedResiduals}
\end{figure}

The test produced the difference in the time domain and the $95\%$ confidence bands shown in Fig. \ref{fig:SimulatedResiduals}. A comparison between the confidence interval and the horizontal axis, representing the null hypothesis of having no difference between the two means, shows that the difference is significant (i.e., the axis is outside of the confidence interval). The difference between the confidence bands and the horizontal axis, the residuals, are plotted in \ref{fig:SimulatedResiduals} below. In Fig. \ref{fig:residualFRF}, the DFT of the residuals is plotted on the frequencies of interest (the ones where the PRTS input is not zero). The representation in the frequency domain allows for a localization of the significant difference between the two groups in the frequency range between $0.5$ and $1$ Hz. 

\begin{figure}[htbp]
	\centering
		\includegraphics[width=1.00\textwidth]{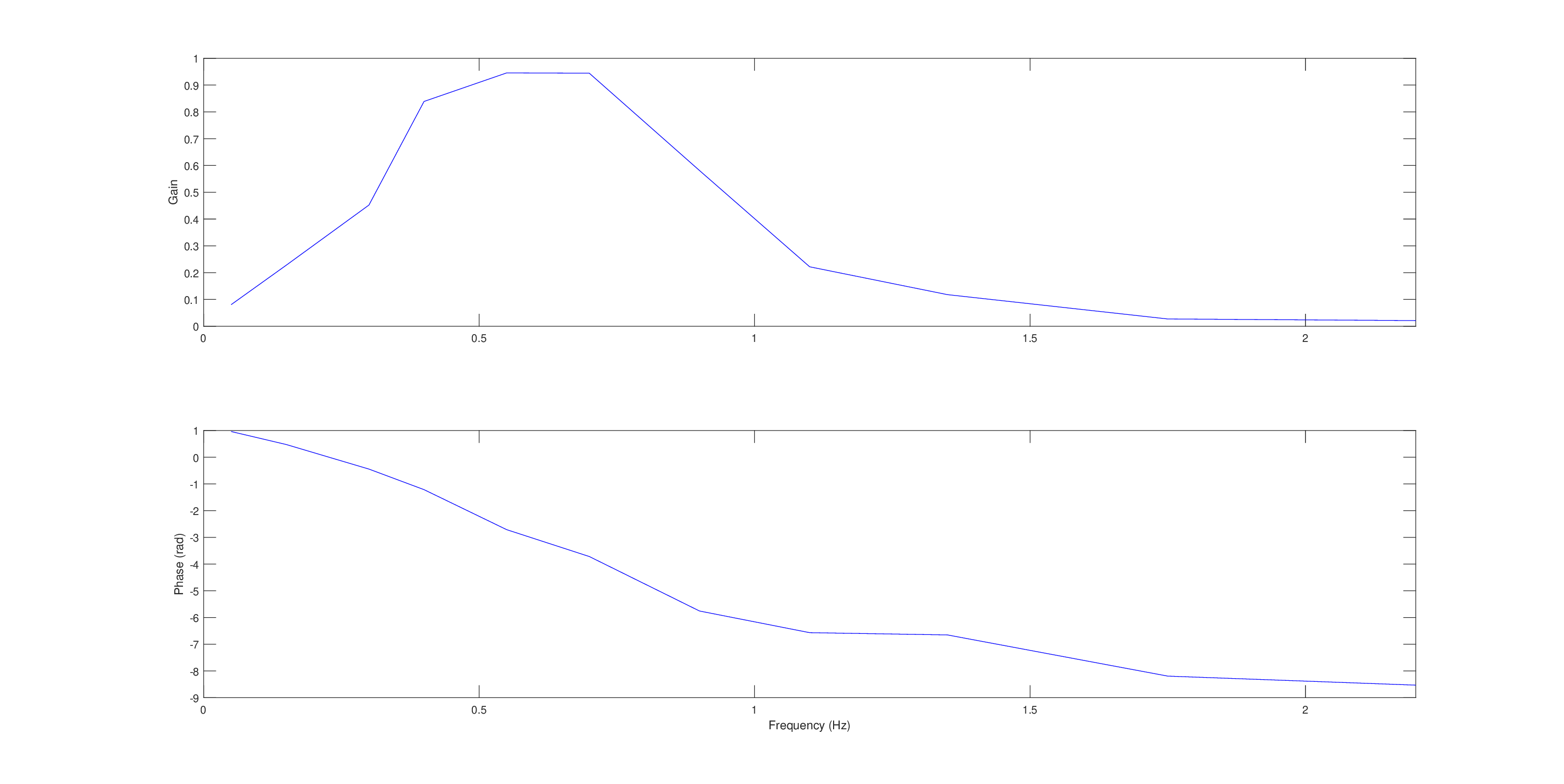}
	\caption{The Fourier transform of the residuals (Fig. \ref{fig:SimulatedResiduals}) shows how the significant difference between the two sample sets is localized between $0.5$ and $1$ Hz.}
	\label{fig:residualFRF}
\end{figure}
\section{Discussion}
\begin{figure}[htbp]
	\centering
		\includegraphics[width=1.00\textwidth]{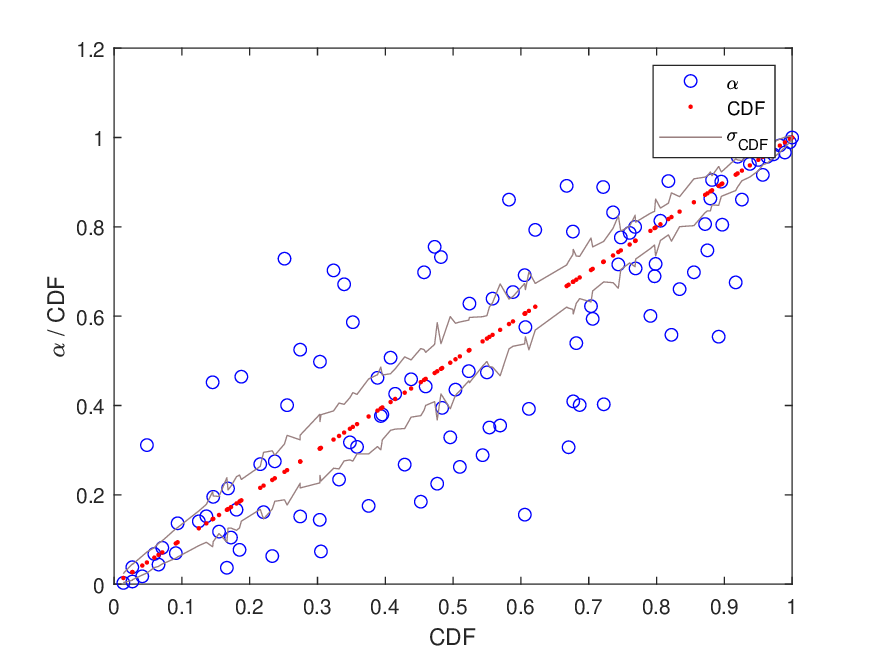}
	\caption{Confidence level $\alpha$ associated with the minimal prediction band and CDF and the STD of the CDF estimated with the bootstrap $\sigma_{CDF}$. The two measures are correlated but different.}
	\label{fig:PointsCDFdiff}
\end{figure}

The minimal prediction bands and the estimated CDF provide a measure of how the sample is close to the mean. The two measures provide a similar value in the example but, in general, are different. In Fig. \ref{fig:PointsCDFdiff}, the confidence level $\alpha$ associated with all the samples is plotted versus the estimated CDF. The graph shows that the two measures appear to be correlated. However, for diagnostics applications, it may be useful to use both the tests in the sample (e.g., considering the sample as anomalous if one of the two measures exceeds a threshold). The estimated pdf can be useful in cases where the distribution is multimodal, where a distance measure between a sample and the average PIR may not represent the sample likelihood.  

The comparison between groups showed how the method could recognize the difference between two groups and, through the Fourier transform of the residuals, it can describe the difference in the frequency domain, the thing that usually is done considering groups of frequencies that are manually selected \cite{joseph2014contribution} or testing each frequency separately\cite{lippi2023human,Akcay2021}.

\section{Conclusions, and Future Work}
The paper presented three new methods to perform statistical tests on FRFs. The minimal prediction bands and the estimation of sample probability estimate how far a sample is from the mean of a distribution with slightly different but comparable results. The test for the difference between unpaired samples provides a way to check the effect of a condition when the subjects are different in the two groups (e.g., comparison between healthy control and test, or placebo and treatment).  

\textbf{Future work} will consist of the application of these measures to quantify the difference between groups of patients (e.g., polyneuropathy) and a control group of healthy subjects and in humanoid robotics assessment to evaluate the human likeness of the produced motion pattern as an alternative to the pseudo-statistics based on the covariance matrix that was proposed in \cite{lippi2020human} with all the advantages discussed in the introduction. The computation of cdf and pdf can be used to evaluate the likelihood of the sample. Strictly speaking, they are based on an arbitrary scalar value (the integral of the distance). This may be useful as a further test to be performed on a sample after comparing it with a prediction band with a fixed $\alpha$ to have a continuous score beside the binary result of such a comparison with the band. Both measures can compare the output of different identified posture control models and decide the best one representing the sample population.
As introduced in the overview, the FRFs are not presented here as a system identification but as the representation of a trial, meaning that the same subject can produce different FRFs under different conditions and inputs. Nevertheless, previous works present examples of system identification using the error on FRF as target function \cite{asslander2015visual,icinco20,icpram21} and identification of posture control as transfer functions \cite{van2005comparison,vanderKooij2007,1400711,pasma2018evidence}. The proposed approach based on Bootstrap can allow for identification techniques based on statistical information (e.g., mean and STD) rather than just on the quadratic error.  

\section*{Appendix: The Code}
The library source code is available in the repository: https://github. com/mcuf-idim/FRF-statistics. Here, the Matlab code of the newly introduced functions is reported for reference.

\subsection{Minimal Prediction Band}
\begin{programcode}{FRF\_MinimalPredictionBand}
\begin{lstlisting}[style=Matlab-editor]
function [avg,sigma,band,Cp,chist,values,alpha] = FRF_MinimalPredictionBand(X,FRFs,phi,sample_time,B)
%[AVG,SIGMA,VOL,CP,CHIST,VALUES,ALPHA] =
%FRF_MINIMALPREDICTIONBAND(X,FRFS,PHI,SAMPLE_TIME,B)
% Computes the minimum Cp that includes the tested FRF X and the empirical
% ALPHA associated with it.
% where avg is average PIR, sigma is the measure of variation of x(t), band
% a two-row matrix with the boundaries of the band. Cp is the threshold
% constant obtained by the bootstrap. FRFS is a matrix where each row
% represents an FRF of the set, phi is the vector of frequencies, and
% SAMPLE_TIME is the sample time of the PIRs. Chist is a vector
% representing the cumulative histogram for the values returned in VALUES.

N=size(FRFs,1); %number of FRFs

sf=1/sample_time;

xt=FRF_pseudoimpulse(X,phi,sf);

ns=length(xt);
yt=zeros(N,ns);
for i=1:N
[x,t]=FRF_pseudoimpulse(FRFs(i,:),phi,sf);
yt(i,:)=x;
end

xm=mean(yt);
sx=std(yt);


STAT=zeros(1,B*N);

s=1;
for b=1:B %GENERATE THE HISTOGRAM
resamp=randi(N,1,N);
yb=yt(resamp,:);
xb=mean(yb);
sb=std(yb);
for n=1:N
db=max(abs(yt(n,:)-xb)./sb);
STAT(s)=db;
s=s+1;
end
end

%% Histogram

STAT=sort(STAT);

[chist, values] = histcounts(STAT,1000,'Normalization','cdf');

[Cp,idx]=max(abs(xt-xm)./sx);

alpha = chist(find(values>Cp,1,'first'));

avg = xm;
sigma = sx;

band=[avg+Cp*sigma;avg-Cp*sigma];
end
\end{lstlisting}
\end{programcode}

\subsection{Estimation of the Probability of a Sample}
\begin{programcode}{FRF\_pdf}
\begin{lstlisting}[style=Matlab-editor]
function [cdf,pdf,sigma_cdf,sigma_pdf] = FRF_pdf(varargin)
% [cdf,pdf,sigma_cdf,sigma_pdf] = FRF_pdf(X,FRFs,phi,sample_time,B,metric)
%
% estimates the cumulative density function and the density function
% associated with the sample X and an STD on their estimation.
% the input METRIC defines the measure used to define the distance
% between X and the mean of the sample FRFs. By default, it is the sum of
% squared residuals
% distance; if METRIC is a function handle, it is applied directly. The
% following strings can be specified:
% - 'squared' sum of squared residuals
% - 'max' maximum difference between two samples

if nargin==5
metric=@(x,y) sum((x-y).^2,2);
elseif nargin==6
metric = varargin{6};
if isa(metric,'function_handle')
disp(''); % so far do nothing and use it straightforward
elseif isa(metric,'string') || isa(metric,'char')
if metric== "squared"
metric=@(x,y) sum((x-y).^2,2);
elseif metric== "max"
metric=@(x,y) max(abs(x-y),[],2);
else
error([metric,' is not a valid metric']);
end
else
error('METRIC must be a string or a function handle');
end
else
error('Input arguments must be 5 or 6');
end
X=varargin{1};
FRFs=varargin{2};
phi=varargin{3};
sample_time=varargin{4};
B=varargin{5};

N=size(FRFs,1); %number of FRFs

Ds=max(1,fix(N/20)); %This will be a parameter in future versions

sf=1/sample_time;

xt=FRF_pseudoimpulse(X,phi,sf);

ns=length(xt); %generalize!

yt=zeros(N,ns);

for i=1:N
[x,t]=FRF_pseudoimpulse(FRFs(i,:),phi,sf);
yt(i,:)=x;
end

xm=mean(yt);
sx=std(yt);

STAT=zeros(1,B);
dSTAT=zeros(1,B);

for b=1:B %GENERATE THE HISTOGRAM
resamp=randi(N,1,N);
yb=yt(resamp,:);
xb=mean(yb);
es=sort(metric(yb,xb));
et=metric(xt,xb);
idx=find(es>et,1,'first');
if isempty(idx)
idx=N;
end
STAT(b)= idx/N;
i1=idx-Ds;
i2=idx+Ds;
if i1<1
i1=1;
i2=1+Ds;
end
if i2>N
i2=N;
i1=N-Ds;
end
dSTAT(b)=(Ds)/(N*(es(i2)-es(i1)));
end

cdf=mean(STAT);
sigma_cdf=std(STAT);

pdf=mean(dSTAT);
sigma_pdf=std(dSTAT);
\end{lstlisting}
\end{programcode}

\subsection{Comparison between unpaired samples}
\label{codeComparison}
\begin{programcode}{FRF\_ConfidenceBandDifference}
\begin{lstlisting}[style=Matlab-editor]
function [avg,sigma,band,Cc,chist,values] = FRF_ConfidenceBandDifference(FRF1,FRF2,phi,sample_time,alpha,B,Bs)
%[AVG,SIGMA,band,CC,CHIST,VALUES] =
%FRF_CONFIDENCEBANDDIFFERENCE(FRFS1,FRFS2,PHI,SAMPLE_TIME,B)
% Confidence bands on the difference between the means of two groups FRF1
% andd FRF2.
%
% B is the number of bootstrap repetitions
% Bs is the number of bootstrap repetitions used to estimate STD
%
% avg is the difference between average PIRs of the groups, sigma is the
% measure of the variation of x(t), and band is a two-row matrix with the boundaries of
% the band. Cp is the threshold constant obtained by the bootstrap. FRFS
% is a matrix where each row represents an FRF of the set, phi is the vector
% of frequencies, and SAMPLE_TIME is the sample time of the PIRs. Chist is
% a vector representing the cumulative histogram for the values returned
% in VALUES.

sf=1/sample_time;

N1=size(FRF1,1); %number of FRFs

x1=FRF_pseudoimpulse(FRF1(1,:),phi,sf);
ns=length(x1);
y1=zeros(N1,ns);
y1(1,:)=x1;

for i=2:N1
[x,t]=FRF_pseudoimpulse(FRF1(i,:),phi,sf);
y1(i,:)=x;
end

N2=size(FRF2,1); %number of FRFs

x1=FRF_pseudoimpulse(FRF2(1,:),phi,sf);
ns=length(x1);
y2=zeros(N1,ns);
y2(1,:)=x1;

for i=2:N2
[x,t]=FRF_pseudoimpulse(FRF2(i,:),phi,sf);
y2(i,:)=x;
end

xm=mean(y1)-mean(y2);

sSTAT=zeros(Bs,ns);
for b=1:Bs
resamp1=randi(N1,1,N1);
resamp2=randi(N2,1,N2);
yb1=y1(resamp1,:);
yb2=y2(resamp2,:);
sSTAT(b,:)=mean(yb1)-mean(yb2);
end
sx=std(sSTAT);

STAT=zeros(1,B);

for b=1:B %GENERATE THE HISTOGRAM
resamp1=randi(N1,1,N1);
resamp2=randi(N2,1,N2);
yb1=y1(resamp1,:);
yb2=y2(resamp2,:);
xb=mean(yb1)-mean(yb2);
for b2=1:Bs
resampb1=randi(N1,1,N1);
resampb2=randi(N2,1,N2);
ybb1=yb1(resampb1,:);
ybb2=yb2(resampb2,:);
sSTAT(b2,:)=mean(ybb1)-mean(ybb2);
end
sb=std(sSTAT);
db=max(abs(xm-xb)./sb);
STAT(b)=db;
end


%% Histogram

STAT=sort(STAT);

[chist, values] = histcounts(STAT,1000,'Normalization','cdf');
Cc=values(find(chist>alpha,1,'first'));

avg=xm;
sigma=sx;

band=[avg+Cc*sigma;avg-Cc*sigma];

end
\end{lstlisting}
\end{programcode}

\bibliographystyle{unsrt}
\bibliography{Bibliography}
\end{document}